\documentclass[12pt,twoside,english,review]{elsarticle}
\usepackage{mathptmx}
\usepackage[T1]{fontenc}
\usepackage[latin9]{inputenc}
\usepackage{geometry}
\usepackage{color}
\usepackage{amsmath}
\usepackage{amssymb}
\usepackage{esint}

\makeatletter
\journal{}

\makeatother

\usepackage{babel}
\begin{document}

\title{On the quasi-streamfunction formalism for waves and vorticity}

\author[UWM]{William Cottrell}

\ead{wcottrell@wisc.edu}

\author[UF]{Miao Tian\corref{cor1}}

\ead{mtian04.18@ufl.edu}

\author[UF]{Alex Sheremet}

\ead{alex@coastal.ufl.edu}

\cortext[cor1]{Corresponding author}

\address[UWM]{Department of Physics, University of Wisconsin, Madison, 1150 University
Avenue, Madison, WI 53706, USA.}

\address[UF]{Department of Civil and Coastal Engineering, University of Florida,
365 Weil Hall, Gainesville, FL 32611, USA.}
\begin{abstract}
The quasi-streamfunction ($\boldsymbol{\Psi}$) formalism proposed
by Kim et. al. (J.W. Kim, K.J. Bai, R.C. Ertekin, W.C. Webster, J.
Eng. Math. 40, 17 (2001)) provides a natural framework for systematically
studying zero-vorticity waves over arbitrary bathymetry. The modified
$\boldsymbol{\Psi}$-formalism developed here discards the original
constraints of zero-vorticity by allowing for vertical vorticity which
is the case of most interest for coastal dynamics. The problem is
reformulated in terms of two dynamical equations on the boundary supplemented
by one equation that represents a kinematic constraint in the interior
of the domain. In this framework, the kinematic constraint can be
solved to express $\boldsymbol{\Psi}$ in terms of the canonically-conjugated
variables $\eta$ and $\phi$. The formalism is demonstrated for horizontally
homogeneous flows over mild topography, where asymptotic formulations
for the Hamiltonian and Lagrangian functions are derived based on
the Helmholz-Hodge decomposition. For potential flows, the asymptotic
form of the Hamiltonian is identical to previous results. The Lagrangian
function is also expressed as an expansion in terms of measurable
variables $\eta$ and $\partial_{t}\eta$, and compared with Zakharov's
formalism where agreement is found for one-dimensional wave scattering.\end{abstract}
\maketitle

\section{Introduction}

A complete understanding of wave evolution requires a consistent and
unified formulation of the interaction of waves moving over varying
topography with non-negligible shearing. Although numerical methods
based on the Navier-Stokes equations are available, a simple lagrangian
formalism that incorporates shearing seems to be lacking. In practice,
one is often restricted to the assumption of gradient flow in order
to approach coastal problems analytically. Another alternative, Clebsch
variables \citep[e.g.,][]{Clebsch_1875,Luke1967}, is a complete and
consistent approach yet is physically unintuitive and not well suited
to the study of wave phenomenon. Shallow water vorticity flows have
also been studied using the shallow-water equation \citep{Kirby1986,MeiLo84};
described as a vortex sheet with constant vertical velocities \citep[e.g., ][]{Evans1975};
as thin shear layers flowing along discrete trenches \citep[e.g., ][]{Kirby1983,Kirby1987};
and using a Hamiltonian formulation proposed by \citep{Shrira1986}
for constant vorticity. However, these studies utilize a diverse variety
of techniques and were targeted towards specific applications while
a unifying formalism for arbitrary forcing and vorticity is still
needed. 

The quasi-streamfunction approach (hereafter denoted the `$\boldsymbol{\Psi}$-formalism')
as studied here was proposed by Kim et al. \citep{Kim01,KimBai03},
and applied to surface waves by \citet{KimBai04} and \citet{ToledoAgnon09}.
(The stream function $\boldsymbol{\Psi}$ was used much earlier by
\citep{Zakharov1999-2d}, in his study of nonlinear interactions of
vortex filaments). Because the $\boldsymbol{\Psi}$-formalism satisfies
exactly the bottom boundary condition, it appears to provide simple
way to systematically describe uneven topography. Although one is
generally limited to the mild-slope approximation in solving the differential
equations generated by the $\Psi$ formalism, the method nevertheless
gives one a convenient means for expanding solutions in terms of a
mild-slope parameter. In addition, the formalism has the advantage
of naturally allowing vertical vorticity; a feature which seems to
have been overlooked in the original formulation \citep{Kim01,KimBai03}. 

The purpose of this study is to remove the zero-vorticity restriction
by incorporating a vertical vorticity. As the largest velocity components
are horizontal in many oceanographic environments, vertical vorticity
typically dominates over horizontal vorticity. Our formalism may be
used, for instance, to study the propagation of waves through sheared
currents or to study vorticity production in wave scattering. We hope
to transform the $\boldsymbol{\Psi}$-formalism into a more useful
tool for the study of wave propagating over inhomogeneous topography
and currents.

Some of the results here have been derived in \citet{Tian_ICCE2012}.
This paper presents an extension to their work. The governing equations
are re-derived in their original Lagrangian formulation Section \ref{sec: governing eqs}
and then solved using an expansion in powers of the free surface displacement
$\eta$ in Section \ref{sec: homogeneous flows}. The derivation also
provides an opportunity to correct a slight abuse of the variational
principle in the original formulation. The Hamiltonian form of the
$\boldsymbol{\Psi}$-formalism is derived in Section \ref{sec: Hamilton Lagrange}.
We propose a Lagrangian only containing measurable quantities $\eta$
and $\partial_{t}\eta$, consistent with Zakharov's one-dimensional
Hamiltonian formalism. The relationship between the $\boldsymbol{\Psi}$-formalism
and the potential flow is investigated in the Appendix, where we propose
an interpretation of the function $\boldsymbol{\Psi}$. Section \ref{sec: sum}
summarizes the results.

\section{Governing Equations \label{sec: governing eqs}}

Throughout this study, $t$ denotes the time and the horizontal vectors
are boldface, e.g., $\mathbf{x}=(x^{1},x^{2})=x^{1}\hat{\mathbf{x}}_{1}+x^{2}\hat{\mathbf{x}}_{2}=x^{j}\hat{\mathbf{x}}_{j}$,
with $j=1,2$. We will prefer Einstein's repeated-index summation
convention (last equality). The origin of the coordinate system is
set on the undisturbed free surface with the vertical axis ($\hat{\mathbf{z}}$)
pointing upward. The hat denotes a unit vector in the direction of
the axis. The free surface is defined by $z=\eta(\mathbf{x},t)$ and
the bottom by $z=-h(\mathbf{x})$. The symbol $\nabla$ denotes the
horizontal gradient.

Define the quasi-streamfunction as a vector: 

\begin{equation}
\boldsymbol{\Psi}(\mathbf{x},z,t)=\int_{-h}^{z}\mathbf{u}(\mathbf{x},s,t)ds.\label{eq: u - Psi}
\end{equation}
The velocity field is defined as, 

\begin{equation}
\mathbf{u}=\boldsymbol{\Psi}_{z},\label{eq: Psi - u}
\end{equation}
where $\mathbf{u}(\mathbf{x},z,t)=u^{j}(\mathbf{x},z,t)\,\hat{\mathbf{x}}_{j}$
is the horizontal velocity vector of components $u^{j}$, with $j=1,2$.
From the continuity of the incompressible fluid, we obtain the vertical
velocity $w$ as: 
\begin{equation}
w=-\nabla\cdot\boldsymbol{\Psi}.\label{eq: Psi - w}
\end{equation}
Defining the total spatial gradient along a given surface $z=\zeta(\mathbf{x},t)$
as 
\begin{equation}
\mathsf{D}_{\zeta}=\nabla\cdot+\left(\nabla\zeta\right)\cdot\partial_{z},\label{eq: D_zeta}
\end{equation}
the total divergence of $\boldsymbol{\Psi}$ on the bottom $z=-h$
is 
\begin{equation}
\mathsf{D}_{-h}\cdot\boldsymbol{\Psi}=-w-(\nabla h)\cdot\mathbf{u}=0,\label{eq: bottom BC}
\end{equation}
which is the standard kinematic bottom boundary condition. This equality
always holds because $\boldsymbol{\Psi}|_{z=-h}=0$ by (\ref{eq: u - Psi}).
Therefore the quasi-streamfunction $\boldsymbol{\Psi}$ unconditionally
satisfies the kinematic bottom boundary condition \citep{KimBai04}.
In fact, one can show that for potential flows the relation between
$\boldsymbol{\Psi}$ and $\boldsymbol{\Phi}$ is similar to the electromagnetic
duality (Appendix).

The dynamics of this system are determined by the Lagrangian density;
see \citep[e.g., ][]{Kim01,KimBai03,KimBai04} 
\begin{equation}
\mathcal{L}=\phi\left[\eta_{t}+\nabla\cdot\boldsymbol{\Psi}+\boldsymbol{\Psi}_{z}\cdot\nabla\eta\right]_{\eta}+\frac{1}{2}\intop_{-h}^{\eta}\left[\left|\boldsymbol{\Psi}_{z}\right|^{2}+\left(\nabla\cdot\boldsymbol{\Psi}\right)^{2}\right]dz-\frac{g}{2}\eta^{2},\label{eq: L_0}
\end{equation}
where $\phi(\mathbf{x},t)$ is a Lagrange multiplier which ensures
that the free-surface kinematic condition is satisfied. The form \eqref{eq: L_0}
can be simplified significantly. Using the total derivative $\mathsf{D}_{\eta}$
\eqref{eq: D_zeta} and the identity (up to total derivatives) $\phi\mathsf{D}_{\eta}\cdot\Psi=-\left(\mathsf{D}_{\eta}\phi\right)\cdot\boldsymbol{\Psi}=-\left(\nabla\phi\right)\cdot\boldsymbol{\Psi},$
the Lagrangian can be written as: 
\begin{equation}
L=\int\mathcal{L}\, d^{2}x;\quad\mathcal{L}=\phi\eta_{t}-\nabla\phi\cdot\boldsymbol{\Psi}+\frac{1}{2}\intop_{-h}^{\eta}\left[\left|\boldsymbol{\Psi}_{z}\right|^{2}+\left(\nabla\cdot\boldsymbol{\Psi}\right)^{2}\right]dz-\frac{g}{2}\eta^{2}.\label{eq: L_1}
\end{equation}
The vertically integrated term of the Lagrangian \eqref{eq: L_1}
may be rewritten as: 
\begin{equation}
L_{vert}=\frac{1}{2}\int\int d^{2}x\intop_{-h}^{\eta}\left[\left|\boldsymbol{\Psi}_{z}\right|^{2}+\left(\nabla\cdot\boldsymbol{\Psi}\right)^{2}\right]dz=\frac{1}{2}\iint d^{2}x\int_{-h}^{\eta}dz\left[\partial_{i}\Theta^{i}-\boldsymbol{\Psi}\cdot\left(\partial_{z}^{2}\boldsymbol{\Psi}+\nabla(\nabla\cdot\boldsymbol{\Psi})\right)\right]\label{eq: L vertical}
\end{equation}

where 
\begin{equation}
\Theta=\boldsymbol{\Psi}\nabla\cdot\boldsymbol{\Psi}+\left(\boldsymbol{\Psi}\cdot\boldsymbol{\Psi}_{z}\right)\hat{\mathbf{z}},\label{eq: Theta 3}
\end{equation}
is a 3-dimensional vector with divergence. To prove the second equality
in (\ref{eq: L vertical}) note that:
\begin{align*}
\partial_{i}\Theta^{i} & =\nabla\cdot\left(\boldsymbol{\Psi}\nabla\cdot\boldsymbol{\Psi}\right)+\left(\boldsymbol{\Psi}\cdot\boldsymbol{\Psi}_{z}\right)_{z}\\
= & \left(\nabla\cdot\Psi\right)^{2}+|\Psi_{z}|^{2}+\Psi\cdot\left(\partial_{z}^{2}\Psi+\nabla\left(\nabla\cdot\Psi\right)\right).
\end{align*}
Because $\Theta=0$ on $z=-h$, applying Gauss's theorem yields 
\[
L_{vert}=\frac{1}{2}\iint_{\eta}\Theta{\color{blue}\cdot}d\mathbf{S}-\frac{1}{2}\iint d^{2}x\int_{-h}^{\eta}\boldsymbol{\Psi}\cdot\left(\partial_{z}^{2}\boldsymbol{\Psi}+\nabla\nabla\cdot\boldsymbol{\Psi}\right)\, dz,
\]
where $d\mathbf{S}=\mathbf{n}dA$, with $dA$ the measure of the area
and $\mathbf{n}$ the normal to the free surface. The first term represents
an integral over the the free surface and the second an integral over
the interior of the fluid. The least action principle 
\[
\delta\int Ldt=0
\]
requires that solutions minimize the action under all possible variations
of the fields. In particular, we may consider variations under which
the surface terms change independently from the interior terms. This
implies that their variations must vanish independently. The variation
of the interior term leads to the ``Laplace''-like equation: 
\begin{equation}
\boldsymbol{\Psi}_{zz}+\nabla(\nabla\cdot\boldsymbol{\Psi})=0\label{eq: psi-Laplace}
\end{equation}
This translates to the statement that the two horizontal components
of vorticity are zero. We impose this condition as a constraint. The
above equation does not fully determine the stream function, $\Psi$,
which must be fixed by specifying the surface values, $\Psi(\eta)$
and $\Psi_{z}(\eta)$. The least action principle demands that we
minimize under all possible configurations of $\Psi$, and, (after
imposing the constraint) these configurations are labeled uniquely
by $\Psi(\eta)$ and $\Psi_{z}\left(\eta\right)$. Now, imposing Eq.
\eqref{eq: psi-Laplace} , and writing

\begin{equation}
d\mathbf{S}=\frac{(-\nabla\eta,1)}{\sqrt{1+|\nabla\eta|{}^{2}}}dA;\;\mbox{with}\; dA=\sqrt{1+|\nabla\eta|{}^{2}}d^{2}x,\label{eq: surface element}
\end{equation}
one finds the following simplified expression for the interior contribution
to the action written entirely in terms of surface functions: 
\begin{equation}
L_{vert}=\frac{1}{2}\iint_{\eta}d^{2}x\,\boldsymbol{\Psi}\cdot\left(\boldsymbol{\Psi}_{z}-\nabla\eta(\nabla\cdot\boldsymbol{\Psi})\right),\label{eq: L vert}
\end{equation}
This yields a very simple expression for the Lagrangian density: 
\begin{equation}
\mathcal{L}=\frac{1}{2}\Psi^{j}\mathbf{K}_{jl}\Psi^{l}-\nabla\phi\cdot\boldsymbol{\Psi}+\phi\eta_{t}-\frac{g}{2}\eta^{2};\;\mbox{with}\;\mathbf{K}_{jl}=(\delta_{jl}\partial_{z}-\partial_{j}\eta\partial_{l}).\label{eq: L_2}
\end{equation}
where $\delta_{jl}$ is the Kronecker symbol. 

One may think of this as a matrix problem with $\Psi$ being a vector,
$\mathbf{K}$ being a matrix, and the integral representing a contraction
of indices. The quantity $\Psi^{j}\mathbf{K}_{jl}\Psi^{l}$ is equivalent,
by definition to $\Psi^{j}\left\{ \mathbf{K}_{jl}\right\} ^{T}\Psi^{l}$,
where $\left\{ \mathbf{K}_{jl}\right\} ^{T}$ is the transpose of
$\mathbf{K}_{jl}$. Since $\Psi^{j}\mathbf{K}_{jl}\Psi^{l}$ and $\Psi^{j}\left\{ \mathbf{K}_{jl}\right\} ^{T}\Psi^{l}$
are equal, we may add them and divide by two. For understanding the
variations of Lagrangian \eqref{eq: L_2} with respect to $\boldsymbol{\Psi}$,
it will be convenient to recast it into a symmetric form by formally
introducing a new operator $\mathbf{T}=\frac{1}{2}\left(\mathbf{K}+\mathbf{K}^{T}\right)$,
where $\mathbf{K}^{T}$ is the transpose of $\mathbf{K}$, i.e., 
\begin{equation}
\mathcal{L}=\frac{1}{2}\Psi^{j}\mathbf{T}_{jl}\Psi^{l}-\nabla\phi\cdot\boldsymbol{\Psi}+\phi\eta_{t}-\frac{g}{2}\eta^{2}.\label{eq: lagrangian}
\end{equation}
The parts of K that are asymmetric under the transpose operation drop
out of the integral after this transform; and T allows us to write
the constraint in terms of a single linear operator. 

When applying the calculus of variation, all functions are varied
independently at each point (ignoring for boundary conditions restricting
the variation for now). Thus, we may consider a variation of the stream
function in one unit of volume completely independently of neighboring
units of volume. Then we may conclude that all solutions must satisfy
the interior equation \eqref{eq: psi-Laplace} without having to solve
all the equations involving the surface terms. After requiring equation
\eqref{eq: psi-Laplace} to hold, we still need to specify $\boldsymbol{\Psi}$
and $\boldsymbol{\Psi}_{z}$ in order to determine a solution uniquely.
Thus, we may think of $\boldsymbol{\Psi}$ and $\boldsymbol{\Psi}_{z}$
as labeling the full solution so that we have just traded variation
over the complete set of functions with variations on a reduced set
of functions which already satisfy the interior equations. The value
of $\boldsymbol{\Psi}$ is then determined by these remaining variations.
The governing equations are given by the variations of the Lagrangian
\eqref{eq: lagrangian} with respect to the variables $\eta,$ $\phi$,
and $\boldsymbol{\Psi}$, respectively: 

\begin{align}
\eta_{t}+D_{\eta}\cdot\boldsymbol{\Psi} & =0\quad\mbox{on}\quad z=\eta\label{eq: phi-eq}\\
\phi_{t}+\nabla\cdot\left(\phi\boldsymbol{\Psi}_{z}\right)-\frac{1}{2}\left[\left|\boldsymbol{\Psi}_{z}\right|^{2}+(\nabla\cdot\boldsymbol{\Psi})^{2}\right]+g\eta & =0\quad\mbox{on}\quad z=\eta\label{eq: eta-eq}\\
\left[\boldsymbol{\Psi}_{z}-(\nabla\cdot\boldsymbol{\Psi})\nabla\eta+\nabla(\boldsymbol{\Psi}\cdot\nabla\eta)-2\nabla\phi+(\nabla\eta\cdot\boldsymbol{\Psi}_{z})\nabla\eta\right]\cdot\delta\boldsymbol{\Psi}\nonumber \\
+\left[\boldsymbol{\Psi}+(\nabla\eta\cdot\boldsymbol{\Psi})\nabla\eta\right]\cdot\delta\boldsymbol{\Psi}_{z} & =0\quad\mbox{on}\quad z=\eta\label{eq: psi-eq}
\end{align}

Note that the variation $\delta\boldsymbol{\Psi}_{z}$ is treated
as independent of $\delta\boldsymbol{\Psi}$ since we have a surface
boundary and total derivatives may not be discarded arbitrarily. (Or,
one may note that both $\Psi$ and $\Psi_{z}$ must be independently
specified in order to determine a solution to the second order equation
\eqref{eq: psi-Laplace}.) Equation \eqref{eq: psi-eq} really is
a vector equation since the two components of $\delta\boldsymbol{\Psi}$
may be varied independently. After applying the variation to $\delta\boldsymbol{\Psi}$
and $\delta\boldsymbol{\Psi}_{z}$, we will obtain two vector equations.

Equations (\ref{eq: phi-eq}-\ref{eq: psi-eq}) are exact conditions
evaluated at the surface that determine $\eta$, $\phi$, and $\boldsymbol{\Psi}$
to arbitrary degree of accuracy. No approximations have been made
so far concerning the interior solutions. Together with the ``Laplace-like''
equation, \eqref{eq: psi-Laplace}, equations (\ref{eq: phi-eq}-\ref{eq: psi-eq})
form the governing equations for the $\boldsymbol{\Psi}$-formalism,
with unknown functions: $\eta(\mathbf{x},t)$, $\phi(\mathbf{x},t)$,
and $\boldsymbol{\Psi}(\mathbf{x},z,t)$.

As will be shown later, $\eta$ and $\phi$ remain canonically-conjugated
variables, which is useful for deriving a Hamiltonian description
of the flow. Equations (\ref{eq: phi-eq}-\ref{eq: eta-eq}) are therefore
dynamical equations for $\eta$ and $\phi$, while \eqref{eq: psi-Laplace}
constrains the vertical structure of the flow. \eqref{eq: psi-eq}
provides a relation between the two dynamic variables via the function
$\boldsymbol{\Psi}$.

The first step toward solving the dynamical equations (\ref{eq: phi-eq}-\ref{eq: eta-eq})
is to eliminate $\boldsymbol{\Psi}$ using the constraint equation
\eqref{eq: psi-eq}. Note that taking the variation of \eqref{eq: lagrangian}
with respect to $\boldsymbol{\Psi}$ gives 
\begin{equation}
\mathbf{T}\boldsymbol{\Psi}-\nabla\phi=0\quad\mbox{on}\quad z=\eta.\label{eq: formal psi-constr}
\end{equation}
with the formal solution 
\begin{equation}
\boldsymbol{\Psi}=\mathbf{T}^{-1}\nabla\phi\quad\mbox{on}\quad z=\eta.\label{eq: formal psi-sol}
\end{equation}

Equation \eqref{eq: formal psi-sol} is a formal representation of
complicated dynamics: $\mathbf{T}$ is defined on the set of 2-dimensional
vector fields $\boldsymbol{\Psi}(\mathbf{x,}z=\eta,t)$ satisfying
the interior equation \eqref{eq: psi-Laplace} and the bottom boundary
condition \eqref{eq: bottom BC}. Inverting $\mathbf{T}$ therefore
means inverting within the image of this space of functions. It may
be shown that $\mathbf{T}$ is invertible on this space and provides
a solution to the constraint. Without further elaboration on this
point (will be presented elsewhere), we note that important information
may be gleamed from equation \eqref{eq: formal psi-sol}: 1) We see
that each term in a perturbative expansion for $\boldsymbol{\Psi}$
will contain only one power of $\phi$ and an arbitrary number of
$\eta$; and 2) the variation with respect to $\boldsymbol{\Psi}$
will produce a total derivative in an effective surface Lagrangian
for $\phi$ and $\eta$. In the present work, we will not pursue the
explicit construction of $\mathbf{T}$, rather, we will work directly
with the equivalent equation \eqref{eq: psi-eq}.

We conclude this section by noting that retaining only the quadratic
terms of the constraint equation \eqref{eq: psi-eq}, the linear theory
of \citep{KimBai03} is retrieved, where 
\[
\boldsymbol{\Psi}_{z}=\nabla\phi+\mbox{higher order terms}.
\]
In the leading order, $\phi$ is equal to the velocity potential at
the surface, although it seems difficult to provide a more intuitive
statement. The physically meaningful variable is the surface elevation
$\eta$; and $\phi$ is simply the variable that is canonically conjugate
to that. In the full theory, this interpretation is corrected by higher
order terms.

\section{Homogeneous flows over slowly varying topography\label{sec: homogeneous flows}}

The formalism presented above provides a consistent means for incorporating
both varying topography and vertical vorticity. We illustrate here
the application of the theory to flows that admits a wave-number Fourier
representation. For simplicity, the discussion will be limited to
slowly varying topography. The analysis of more complicated settings
will be presented elsewhere.

\subsection{Interior solutions }

Assuming that the problem is horizontally homogeneous, the unknown
functions admit wave number Fourier representation 
\begin{equation}
\left(\begin{array}{c}
\eta\\
\phi\\
\boldsymbol{\Psi}
\end{array}\right)=\int\frac{d^{2}k}{2\pi}\,\left(\begin{array}{c}
\eta_{\mathbf{k}}\\
\phi_{\mathbf{k}}\\
\boldsymbol{\Psi}_{\mathbf{k}}
\end{array}\right)e^{i\mathbf{k}\cdot\mathbf{x}};\quad\left(\begin{array}{c}
\eta_{\mathbf{k}}\\
\phi_{\mathbf{k}}\\
\boldsymbol{\Psi}_{\mathbf{k}}
\end{array}\right)=\int d^{2}x\left(\begin{array}{c}
\eta\\
\phi\\
\boldsymbol{\Psi}
\end{array}\right)e^{-i\mathbf{k}\cdot\mathbf{x}}\label{eq: FT}
\end{equation}
where $\mathbf{k}=k^{j}\hat{\mathbf{x}}_{j}$ is the wave number vector,
and the $g_{\mathbf{k}}=[g]_{\mathbf{k}}$ is the Fourier transform
of $g$. The short-hand notation $[\cdots]_{\mathbf{k}}$ will later
simplify the handling of convolution products resulting from the Fourier
transform of nonlinear terms. Because the functions $\eta$ and $\phi$
are real, their transforms satisfy the regular symmetry conditions,
e.g., $\eta_{\mathbf{k}}=\eta_{-\mathbf{k}}^{*}$ with the asterisk
denoting the complex conjugate. In Fourier space, the Helmholtz-Hodge
decomposition \citep[e.g., ][]{ChorinMarsden93} of the flow in terms
of $\Psi$ is 
\begin{equation}
\boldsymbol{\Psi}=\boldsymbol{\theta}+\boldsymbol{\Gamma};\quad\left(\begin{array}{c}
\boldsymbol{\theta}\\
\boldsymbol{\Gamma}
\end{array}\right)=\int\frac{d^{2}k}{2\pi}\,\left(\begin{array}{c}
\theta_{\mathbf{k}}\hat{\mathbf{k}}\\
\Gamma_{\mathbf{k}}\,\hat{\mathbf{z}}\times\hat{\mathbf{k}}
\end{array}\right)e^{i\mathbf{k}\cdot\mathbf{x}};\label{eq: Psi L-T}
\end{equation}
where $\boldsymbol{\theta}$ and $\boldsymbol{\Gamma}$ are longitudinal
and transversal components representing the curl-free and the divergence-free
motions. It may be checked that $\nabla\times\boldsymbol{\theta}=\nabla\cdot\boldsymbol{\Gamma}=0$,
and also $\mbox{curl}_{3}\mathbf{u}_{\theta}=\mbox{div}_{3}\mathbf{u}_{\boldsymbol{\Gamma}}=0$,
where $\mathbf{u}_{\theta,\gamma}$ are the velocity components defined
through equation \eqref{eq: Psi - u}, and $\mbox{curl}_{3}$ and
$\mbox{div}_{3}$ are the 3-dimensional versions of the operators.
Both components satisfy the required Fourier symmetries for real physical
domain functions. Therefore we obtain $\nabla(\nabla\cdot\boldsymbol{\Gamma})=0$
and $\nabla(\nabla\cdot\boldsymbol{\theta})=\int\frac{d^{2}k}{2\pi}\left(-k^{2}\theta_{\mathbf{k}}e^{i\mathbf{k}\cdot\mathbf{x}}\right)$.
Substituting decomposition \eqref{eq: Psi L-T} into the governing
equation \eqref{eq: psi-Laplace} for the interior flow yields 
\begin{equation}
\partial_{z}^{2}\theta_{\mathbf{k}}-k^{2}\theta_{\mathbf{k}}=0;\quad\partial_{z}^{2}\Gamma_{\mathbf{k}}=0,\label{eq: Laplace L-T}
\end{equation}
with $k$ ( $k^{2}=k^{j}k^{j}$) the absolute value of the wave number.
In the physical domain the first equation is simply the Laplace equation
for the curl-free component of the flow. For mildly sloping bottoms
(e.g., \citep{Mei-book}), the solution of the equation for $\theta_{\mathbf{k}}$
is the usual 
\begin{equation}
\theta_{\mathbf{k}}=\frac{\sinh\left[k(z+h)\right]}{\sinh(kh)}\,\vartheta_{\mathbf{k}},\label{eq: wkb vert}
\end{equation}
where the wave number is assumed to be a slowly varying function of
the horizontal coordinate. The second equation produces trivial linear
solutions, suitable for describing sheared currents. Explicitly, we
shall take $\Gamma_{\mathbf{k}}$ to be of the form: 
\begin{equation}
\Gamma_{\mathbf{k}}=\frac{z+h'}{h'}\gamma_{\mathbf{k}}.\label{eq:wkb_Gamma}
\end{equation}

Note that the actual velocity is obtained by taking a z derivative
of the streamfunction. Thus, a linear term in $\Gamma$ translates
to a velocity that is constant in depth as required.

Impermeability is automatically satisfied in this formalism so long
as $\boldsymbol{\Psi}\left(z=-h\right)=0$, as may be seen by equation
\eqref{eq: bottom BC}. Since equations \eqref{eq: FT}-\eqref{eq:wkb_Gamma}
are consistent with $\boldsymbol{\Psi}\left(z=-h\right)=0$, there
is no violation of the impermeability condition. Here, we use an $h'$
rather than an $h$ to leave open the possibility that the flow represented
by $\Gamma$ terminates at some finite depth which is not necessarily
the actual bottom. (This phenomenon is observed in surface currents
in the open ocean.) For simplicity, we shall henceforth assume that
$h'=h$. 

The vertical vorticity is an unambiguous physical quantity and it
must be specified as an input or initial condition for the particular
application in mind. For any choice of initial velocity profile one
may solve for the relevant $\Gamma_{\mathbf{k}}$ by inverting equations
\eqref{eq: Psi L-T} and \eqref{eq:wkb_Gamma}. The dynamical equations
then specify the future evolution of the system. For the readers convenience
we invert (\ref{eq: Psi L-T}) in order to provide the expression
for $\gamma_{\mathbf{k}}$ in terms of the observed or modeled velocity
profile.

\begin{equation}
h'\int d^{2}\mathbf{x}\left(\hat{\mathbf{z}}\times\hat{\mathbf{k}}\right)\cdot\mathbf{u}\, e^{-i\mathbf{k}\cdot\mathbf{x}}=\gamma_{\mathbf{k}}
\end{equation}

\subsection{Perturbation solution}

A common approach to seek solutions for surface-gravity wave equations
such as, \eqref{eq: psi-Laplace}, and (\ref{eq: phi-eq}-\ref{eq: formal psi-constr}),
is to expand them in powers of $\eta$. The governing equations may
either be expanded directly, or equivalently, re-derived from the
expanded Lagrangian. Keeping up to cubic terms in the unknown functions
yields the following expansions

\begin{align}
\eta_{t}+\nabla\cdot\left(\boldsymbol{\Psi}+\eta\boldsymbol{\Psi}_{z}+\frac{1}{2}\eta^{2}\boldsymbol{\Psi}_{zz}\right) & =0\;\mbox{on}\; z=0,\label{eq: phi-eq-exp}\\
\phi_{t}+\nabla\phi\cdot\boldsymbol{\Psi}_{z}+\eta\nabla\phi\cdot\boldsymbol{\Psi}_{zz}-\frac{1}{2}\left[\left|\boldsymbol{\Psi}_{z}\right|^{2}+(\nabla\cdot\boldsymbol{\Psi})^{2}\right]\nonumber \\
-\frac{1}{2}\eta\left[\left(\boldsymbol{\Psi}_{z}\right)^{2}+(\nabla\cdot\boldsymbol{\Psi})^{2}\right]_{z}+g\eta & =0\;\mbox{on}\; z=0,\label{eq:eta-eq-exp}\\
\left[\boldsymbol{\Psi}_{z}-\nabla\phi-\nabla(\eta\nabla\cdot\boldsymbol{\Psi})-\frac{1}{2}\nabla\left(\eta^{2}\nabla\cdot\boldsymbol{\Psi}_{z}\right)\right]\delta\boldsymbol{\Psi}+\eta\left\{ \boldsymbol{\Psi}_{z}-\nabla\phi\right.\nonumber \\
\left.-\nabla\eta(\nabla\cdot\boldsymbol{\Psi})-\eta\left[\nabla(\nabla\cdot\boldsymbol{\Psi})\right]\right\} \delta\boldsymbol{\Psi}_{z}+\frac{1}{2}\eta^{2}\left(\boldsymbol{\Psi}_{z}-\nabla\phi\right)\delta\boldsymbol{\Psi}_{zz} & =0,\;\mbox{on}\; z=0\label{eq:psi-eq-exp}
\end{align}

where terms containing fourth- (or higher) order products of $\eta$,
$\phi$, and $\boldsymbol{\Psi}$ have been neglected.

A solution for $\boldsymbol{\Psi}$ in terms of $\eta$ and $\phi$
can be obtained by substituting the form given by equation \eqref{eq: wkb vert}
into the constraint equation \eqref{eq:psi-eq-exp} and considering
variations of $\vartheta$ and $\gamma$ separately. After some algebra,
separating the curl-free and divergence-free components, and neglecting
terms of order higher than cubic, the constraint equation \eqref{eq:psi-eq-exp}
yields 

\begin{align}
\vartheta_{\mathbf{k}}+\int\frac{dkdk_{1}}{2\pi}\,\left[\left(m_{11}\right)_{\mathbf{k}-\mathbf{k}_{1}}\vartheta_{\mathbf{k}_{1}}+\left(m_{12}\right)_{\mathbf{k}-\mathbf{k}_{1}}\gamma_{\mathbf{k}_{1}}\right] & =F_{1},\label{eq: psi-th-perturb}\\
\gamma_{\mathbf{k}}+\int\frac{dkdk_{1}}{2\pi}\left[\left(m_{21}\right)_{\mathbf{k}-\mathbf{k}_{1}}\vartheta_{\mathbf{k}_{1}}+\left(m_{22}\right)_{\mathbf{k}-\mathbf{k}_{1}}\gamma_{\mathbf{k}_{1}}\right] & =F_{2},\label{eq: psi-gm-perturb}
\end{align}

where $\mathbf{k}$, $\mathbf{k}_{1}$ are wave numbers. We use the
following short-hand conventions: $[...]_{\mathbf{k}}$ for the Fourier
transform \eqref{eq: FT}; $\mbox{th}kh$ for $\tanh kh$; and $\mbox{cth}kh$
for $\coth kh$. The coefficients of the left-hand side are 

\begin{align}
\left(m_{11}\right)_{\mathbf{k},\mathbf{k}_{1}} & =\left[\mathbf{k}_{1}\cdot\hat{\mathbf{k}}\mbox{cth}\left(k_{1}h\right)+k_{1}\mbox{th}\left(k_{1}h\right)\right]\eta_{\mathbf{k}_{1}}\nonumber \\
+ & \frac{1}{2}\left[k_{1}\mathbf{k}_{1}\cdot\hat{\mathbf{k}}+kk_{1}+\mbox{th}\left(kh\right)\,\mbox{cth}\left(k_{1}h\right)\,\left(\mathbf{k}_{1}\cdot\mathbf{k}+k_{1}^{2}\right)\right][\eta^{2}]_{\mathbf{k}-\mathbf{k}_{1}}\nonumber \\
\left(m_{12}\right)_{\mathbf{k},\mathbf{k}_{1}} & =\frac{1}{h}\left(\hat{\mathbf{z}}\times\hat{\mathbf{k}}_{1}\right)\cdot\hat{\mathbf{k}}\,\eta_{\mathbf{k}-\mathbf{k}_{1}}+\frac{1}{2h}k\,\mbox{th}\left(kh\right)\,\left(\hat{\mathbf{z}}\times\hat{\mathbf{k}}_{1}\right)\cdot\hat{\mathbf{k}}[\eta^{2}]_{\mathbf{k}-\mathbf{k}_{1}}\nonumber \\
\left(m_{21}\right)_{\mathbf{k},\mathbf{k}_{1}} & =\mathbf{k}_{1}\cdot(\hat{\mathbf{z}}\times\hat{\mathbf{k}})\,\mbox{cth}\left(k_{1}h\right)\,\eta_{\mathbf{k}-\mathbf{k}_{1}}+\frac{1}{2}k_{1}\mathbf{k}_{1}\cdot\left(\hat{\mathbf{z}}\times\hat{\mathbf{k}}\right)\,[\eta^{2}]_{\mathbf{k}-\mathbf{k}_{1}}\nonumber \\
\left(m_{22}\right)_{\mathbf{k},\mathbf{k}_{1}} & =\frac{1}{h}\hat{\mathbf{k}}_{1}\cdot\hat{\mathbf{k}}\,\eta_{\mathbf{k}-\mathbf{k}_{1}}\label{eq:qq}
\end{align}

and the right-hand side terms are 

\begin{align}
F_{1} & =i\,\left(\mbox{th}kh\right)\phi_{\mathbf{k}}+i\int\frac{dkdk_{1}}{2\pi}\mathbf{k}_{1}\cdot\hat{\mathbf{k}}\eta_{\mathbf{k}-\mathbf{k}_{1}}\phi_{\mathbf{k}_{1}}+\frac{i}{2}\int\frac{dkdk_{1}}{2\pi}\,\mathbf{k}_{1}\cdot\mathbf{k}\mbox{th}kh[\eta^{2}]_{\mathbf{k}-\mathbf{k}_{1}}\phi_{\mathbf{k}_{1}},\nonumber \\
F_{2} & =i\iint\frac{dkdk_{1}}{2\pi}k_{1}\cdot(\hat{\mathbf{z}}\times\hat{\mathbf{k}})\eta_{\mathbf{k}-\mathbf{k}_{1}}\phi_{\mathbf{k}_{1}}.\label{eq: psi-th-gm-Fj}
\end{align}

Equations (\ref{eq: psi-th-perturb}-\ref{eq: psi-gm-perturb}) are
linear in $\vartheta$ and $\gamma$ to any order in the nonlinearity
and may be solved in terms of $\phi$ and $\eta$. In a matrix form,
the system (\ref{eq: psi-th-perturb}-\ref{eq: psi-gm-perturb}) is
\begin{equation}
\left(I+M\right)\psi=F,\;\mbox{with},\;\psi=\left(\begin{array}{c}
\vartheta\\
\gamma
\end{array}\right),\; M_{jl}\psi^{j}=\int\frac{dkdk_{1}}{2\pi}\left(m_{jl}\right)_{\mathbf{k},\mathbf{k}_{1}}\left(\psi^{j}\right)_{\mathbf{k}_{1}},\label{eq: psi-th-gm formal}
\end{equation}
where $I$ is the identity matrix and $j,l=1,2$. Because the elements
of the matrix $M$ are higher ordered, equation \eqref{eq: psi-th-gm formal}
may be inverted directly to the order of accuracy required as 
\begin{equation}
\psi=(I-M+M^{2}-...)F\label{eq: psi-th-gm sol formal}
\end{equation}
The procedure to solve for $\vartheta$ and $\gamma$ is now straightforward,
albeit tedious. After some algebra, one finds the following solutions
up to third order terms: 

\begin{align}
\vartheta_{\mathbf{k}} & =i\mbox{th}\left(kh\right)\,\phi_{\mathbf{k}}-i\int\frac{dkdk_{1}}{2\pi}\, k_{1}\mbox{th}\left(kh\right)\,\mbox{th}k_{1}h\,\eta_{\mathbf{k}-\mathbf{k}_{1}}\phi_{\mathbf{k}_{1}}\nonumber \\
 & +i\int\frac{dkdk_{1}dk_{2}}{\left(2\pi\right)^{2}}\, W_{\mathbf{k}\mathbf{k}_{1}\mathbf{k}_{2}}\,\eta_{\mathbf{k}-\mathbf{k}_{1}}\eta_{\mathbf{k}_{1}-\mathbf{k}_{2}}\phi_{\mathbf{k}_{2}},\label{eq:psi-th sol}\\
\gamma_{\mathbf{k}} & =i\frac{dkdk_{1}dk_{2}}{\left(2\pi\right)^{2}}\, k_{1}\mbox{th}\left(k_{1}h\right)\,\left(\mathbf{k}-\frac{1}{2}\mathbf{k}_{1}\right)\cdot(\hat{\mathbf{z}}\times\hat{\mathbf{k}})\eta_{\mathbf{k}-\mathbf{k}_{1}}\eta_{\mathbf{k}_{1}-\mathbf{k}_{2}}\phi_{\mathbf{k}_{2}}\label{eq: psi-gm sol}
\end{align}

with the coefficient 

\begin{align}
W_{\mathbf{k}\mathbf{k}_{1}\mathbf{k}_{2}} & =k_{2}\left[\mathbf{k}_{1}\cdot\hat{\mathbf{k}}+k\mbox{th}\left(kh\right)\,\mbox{th}\left(k_{1}h\right)\right]\mbox{th}\left(k_{2}h\right)\nonumber \\
 & -\frac{1}{2}\left(k_{2}\mathbf{k}_{2}\cdot\hat{\mathbf{k}}+kk_{2}\right)\mbox{th}\left(k_{2}h\right)-\frac{1}{2}k^{2}\mbox{th}\left(kh\right).\label{eq:T coeff}
\end{align}

Remarkably, the inclusion of shear does not affect the solution for
$\vartheta$ to this order. Furthermore, it will become apparent that
$\gamma$ itself does not contribute to 4-wave interactions, since
its contribution is eliminated by vector identities at this order.
Thus, shear does not affect 3- and 4-wave interactions in an isotropic
background. $\gamma$ becomes significant with the inclusion of a
strong shear as a background forcing in the Lagrangian, which will
be pursued in a future work.

\section{Hamiltonian and Lagrangian formalisms\label{sec: Hamilton Lagrange} }

Within the assumptions made in the development of the approach presented
here, it is possible to derive Hamiltonian and Lagrangian formalism.
These allow for direct comparison with existing Hamiltonian theories
and will provide a simpler basis for applications, based on observable
quantities.

\subsection{Hamiltonian formulation}

The derivation of a Hamiltonian formalism starts by noting that, as
in the potential formulation for the linear problem, $\eta$ and $\phi$
are canonical variables. For the $\boldsymbol{\Psi}$-formalism, equation
\eqref{eq: L_0} shows that $\frac{dL}{d\left(\eta_{t}\right)}=\phi$.
The only question is whether or not there is some additional \textquoteleft{}hidden\textquoteright{}
dependence on $\eta_{t}$ via $\boldsymbol{\Psi}$. However, the explicit
equations for $\boldsymbol{\Psi}$ (e.g., constraint \eqref{eq: psi-eq})
show that $\boldsymbol{\Psi}$ is independent on $\eta_{t}$. Thus,
the usual argument that $\phi$ and $\eta$ are canonically-conjugated
variables follows in this formalism as well.

Indeed, the conjugate momentum of the dynamical variable $\eta$ is
\begin{equation}
\phi=\frac{\partial\mathcal{L}}{\partial\eta_{t}},\label{eq: phi - momentum}
\end{equation}
as can be seen from the definition \eqref{eq: lagrangian} of the
Lagrangian, and noting that $\boldsymbol{\Psi}$ has no explicit dependence
on $\eta_{t}$ (e.g., solution \eqref{eq: formal psi-sol}). The Legendre
transformation 
\begin{equation}
H=\int d^{2}x\,\mathcal{H}=\int d^{2}x\left(\phi\eta_{t}-\mathcal{L}\right)\label{eq: Legendre}
\end{equation}
then yields 
\begin{equation}
\mathcal{H}=\nabla\phi\cdot\boldsymbol{\Psi}-\frac{1}{2}\boldsymbol{\Psi}\cdot\left(\boldsymbol{\Psi}_{z}-\nabla\eta(\nabla\cdot\boldsymbol{\Psi})\right)+\frac{g}{2}\eta^{2}\label{eq: hamiltonian}
\end{equation}
where $\boldsymbol{\Psi}$ is given by equation \eqref{eq: formal psi-sol},
written in terms of $\phi$ and $\eta$. An explicit form for the
Hamiltonian is obtained by substituting equations \eqref{eq:psi-th sol}
and \eqref{eq: psi-gm sol} into \eqref{eq: hamiltonian}. As mentioned
before, \eqref{eq: psi-gm sol} does not contribute at this order.
Using the symmetric form of convolution products, the results in \citep{Zakharov99}
are retrieved exactly to order $O\left(\epsilon^{5}\right)$ 
\begin{align}
H & =H_{2}+H_{3}+H_{4}+O\left(\epsilon^{5}\right),\label{eq: H expansion}\\
H_{2} & =\frac{1}{2}\int dk\left(k\,\mbox{th}\left(kh\right)\left|\phi_{\mathbf{k}}\right|^{2}+g\left|\eta_{\mathbf{k}}\right|^{2}\right),\nonumber \\
H_{3} & =\frac{1}{2}\int\frac{dk_{1}dk_{2}dk_{3}}{2\pi}\; T_{\mathbf{k}_{1}\mathbf{k}_{2}}^{(1)}\phi_{\mathbf{k}_{1}}\phi_{\mathbf{k}_{2}}\eta_{\mathbf{k}_{3}}\delta\left(\mathbf{k}_{1}+\mathbf{k}_{2}+\mathbf{k}_{3}\right),\nonumber \\
H_{4} & =\frac{1}{2}\int\frac{dk_{1}dk_{2}dk_{3}dk_{4}}{4\pi^{2}}T_{\mathbf{\mathbf{k}_{1}\mathbf{k}_{2}\mathbf{k}_{3}}\mathbf{k}_{4}}^{\left(2\right)}\phi_{\mathbf{k}_{1}}\phi_{\mathbf{k}_{2}}\eta_{\mathbf{k}_{3}}\eta_{\mathbf{k}_{4}}\delta\left(\mathbf{k}_{1}+\mathbf{k}_{2}+\mathbf{k}_{3}+\mathbf{k}_{4}\right),\nonumber 
\end{align}

where $\delta$ is the Dirac delta, and the interaction coefficients
in symmetric form are 

\begin{align}
T_{\mathbf{k}_{1}\mathbf{k}_{2}}^{(1)} & =-\mathbf{k}_{1}\cdot\mathbf{k}_{2}-\left|\mathbf{k}_{1}\right|\left|\mathbf{k}_{2}\right|\mbox{th}\left(k_{1}h\right)\,\mbox{th}\left(k_{2}h\right),\label{eq:H coeff}\\
T_{\mathbf{k}_{1}\mathbf{k}_{2}\mathbf{k}_{3}\mathbf{k}_{4}}^{(2)} & =-\frac{1}{2}k_{1}k_{2}^{2}\mbox{th}\left(k_{1}h\right)-\frac{1}{2}k_{2}k_{1}^{2}\mbox{th}\left(k_{2}h\right)\nonumber \\
 & +\frac{1}{4}k_{1}k_{2}\left|\mathbf{k}_{1}+\mathbf{k}_{3}\right|\mbox{th}\left(k_{1}h\right)\,\mbox{th}\left(k_{2}h\right)\,\mbox{th}\left(\left|\mathbf{k}_{1}+\mathbf{k}_{3}\right|h\right)\nonumber \\
 & +\frac{1}{4}k_{1}k_{2}\left|\mathbf{k}_{2}+\mathbf{k}_{3}\right|\mbox{th}\left(k_{1}h\right)\,\mbox{th}\left(k_{2}h\right)\mbox{th}\left(\left|\mathbf{k}_{2}+\mathbf{k}_{3}\right|h\right)\nonumber \\
 & +\frac{1}{4}k_{1}k_{2}\left|\mathbf{k}_{1}+\mathbf{k}_{4}\right|\mbox{th}\left(k_{1}h\right)\,\mbox{th}\left(k_{2}h\right)\,\mbox{th}\left(\left|\mathbf{k}_{1}+\mathbf{k}_{4}\right|h\right)\nonumber \\
 & +\frac{1}{4}k_{1}k_{2}\left|\mathbf{k}_{2}+\mathbf{k}_{4}\right|\mbox{th}\left(k_{1}h\right)\mbox{th}\left(k_{2}h\right)\mbox{th}\left(\left|\mathbf{k}_{2}+\mathbf{k}_{4}\right|h\right).\nonumber 
\end{align}

\subsection{Lagrangian formulation}

Based on the expansion described above, we are seeking here a Lagrangian
description based on observable quantities, i.e., the generalized
coordinate $\eta$ and generalized velocity $\partial_{t}\eta$. The
first step is to eliminate $\phi$ by solving its equation of motion
\eqref{eq: phi-eq} for $\boldsymbol{\Psi}$ in terms of $\eta$ and
$\partial_{t}\eta$. The fact that $\gamma$ will not contribute at
quartic order gives us the freedom to ignore $\gamma$ and just solve
for $\vartheta$. Note that this does not preclude, from including
background vorticity in the solution of $\vartheta_{\mathbf{k}}$.
Following a similar procedure as before yields 

\begin{align}
\vartheta_{\mathbf{k}} & =\frac{i}{k}\left(\partial_{t}\eta_{\mathbf{k}}\right)-i\int\frac{dkdk_{1}}{2\pi}\left[\hat{\mathbf{k}}\cdot\hat{\mathbf{k}}_{1}\,\mbox{cth}\left(k_{1}h\right)\right]\,\eta_{\mathbf{k}-\mathbf{k_{1}}}\left(\partial_{t}\eta_{\mathbf{k}_{1}}\right)\nonumber \\
 & +i\int\frac{dkdk_{1}dk_{2}}{\left(2\pi\right)^{2}}V_{\mathbf{k}\mathbf{k}_{1}\mathbf{k}_{2}}\eta_{\mathbf{k}-\mathbf{k}_{1}}\eta_{\mathbf{k}_{1}-\mathbf{k_{2}}}\left(\partial_{t}\eta_{\mathbf{k}_{2}}\right)
\end{align}

with the coefficient 

\begin{equation}
V_{\mathbf{k}\mathbf{k}_{1}\mathbf{k}_{2}}=\left(\hat{\mathbf{k}}\cdot\hat{\mathbf{k}}_{1}\right)\left(\mathbf{k}_{1}\cdot\hat{\mathbf{k}}_{2}\right)\mbox{cth}\left(k_{1}h\right)\,\mbox{cth}\left(k_{2}h\right)\,-\frac{1}{2}\hat{\mathbf{k}}\cdot\mathbf{k}_{2}.\label{eq: V coeff}
\end{equation}

Substituting this back into equation \eqref{eq: L_2} (via \eqref{eq: wkb vert},
and \eqref{eq: FT}) and ignoring the $\phi$ constraint, the Lagrangian,
valid up to quartic order, reads

\noindent 
\begin{align}
L & =L_{2}+L_{3}+L_{4}+O\left(\epsilon^{5}\right),\label{eq:L(eta,eta_t) terms}\\
L_{2} & =\frac{1}{2}\int d^{2}k\left[\frac{\mbox{cth}\left(kh\right)}{k}|\partial_{t}\eta_{\mathbf{k}}|^{2}-\frac{g}{2}|\eta_{\mathbf{k}}|^{2}\right]\nonumber \\
L_{3} & =\frac{1}{2}\int\frac{dk_{1}dk_{2}dk_{3}}{2\pi}G_{\mathbf{\mathbf{k}_{1}\mathbf{k}_{2}\mathbf{k}_{3}}\mathbf{k}_{4}}^{\left(1\right)}\left(\partial_{t}\eta_{\mathbf{k}_{1}}\right)\left(\partial_{t}\eta_{\mathbf{k}_{2}}\right)\eta_{\mathbf{k}_{3}}\delta\left(\mathbf{k}_{1}+\mathbf{k}_{2}+\mathbf{k}_{3}\right)\nonumber \\
L_{4} & =\frac{1}{2}\int\frac{dk_{1}dk_{2}dk_{3}dk_{4}}{4\pi^{2}}G_{\mathbf{\mathbf{k}_{1}\mathbf{k}_{2}\mathbf{k}_{3}}\mathbf{k}_{4}}^{\left(2\right)}\left(\partial_{t}\eta_{\mathbf{k}_{3}}\right)\left(\partial_{t}\eta_{\mathbf{k}_{4}}\right)\eta_{\mathbf{k}_{1}}\eta_{\mathbf{k}_{2}}\delta\left(\mathbf{k}_{1}+\mathbf{k}_{2}+\mathbf{k}_{3}+\mathbf{k}_{4}\right)\nonumber 
\end{align}
with the interaction coefficients 

\noindent 
\begin{align}
G_{\mathbf{\mathbf{k}_{1}\mathbf{k}_{2}\mathbf{k}_{3}}\mathbf{k}_{4}}^{\left(1\right)} & =\left(1+\hat{\mathbf{k}}_{1}\cdot\hat{\mathbf{k}}_{2}\right)\,\mbox{cth}(k_{1}h)\,\mbox{cth}\left(k_{2}h\right)\label{eq: L_3 interaction coeff}\\
G_{\mathbf{\mathbf{k}_{1}\mathbf{k}_{2}\mathbf{k}_{3}}\mathbf{k}_{4}}^{\left(2\right)} & =\mbox{cth}\left(k_{1}h\right)\left\{ \left[\left|\mathbf{k}_{1}\right|-\hat{\mathbf{k}}_{1}\cdot\mathbf{k}_{2}+2\left|\mathbf{k}_{1}+\mathbf{k}_{4}\right|\left(\frac{\mathbf{k}_{1}+\mathbf{k}_{4}}{|\mathbf{k}_{1}+\mathbf{k}_{4}|}\right)\cdot\hat{\mathbf{k}}_{1}\right]\,\right.\nonumber \\
 & +\left.\left[\left(\frac{\mathbf{k}_{1}+\mathbf{k}_{4}}{|\mathbf{k}_{1}+\mathbf{k}_{4}|}\right)\cdot\hat{\mathbf{k}}_{1}\right]\left[\left(\frac{\mathbf{k}_{1}+\mathbf{k}_{4}}{|\mathbf{k}_{1}+\mathbf{k}_{4}|}\right)\cdot\hat{\mathbf{k}}_{2}\right]\left|\mathbf{k}_{1}+\mathbf{k}_{4}\right|\,\mbox{cth}\left(k_{2}h\right)\,\mbox{cth}\left(|\mathbf{k}_{1}+\mathbf{k}_{4}|h\right)\right\} .\label{eq: L_4 interaction coeff}
\end{align}

\noindent One could make it symmetric by switching indices. It is
easy to check that $L_{2}$ is consistent with the results given by
\citet{Zakharov1999-2d} via performing a Legendre transform of \eqref{eq: H expansion}.
$L_{3}$ also agrees with Zakharov's formalism if we substitute the
first order relationship between $\partial_{t}\eta$ and $\phi$ into
\eqref{eq:H coeff}. Following the same procedure, $L_{4}$ shows
consistency to \eqref{eq:H coeff} in 1-D situations (e.g., $\hat{\mathbf{k}}_{1},\:\hat{\mathbf{k}}_{4}$
and $\hat{\mathbf{k}}_{2},\:\hat{\mathbf{k}}_{3}$ are in the opposite
direction). The dot products in $G_{\mathbf{\mathbf{k}_{1}\mathbf{k}_{2}\mathbf{k}_{3}}\mathbf{k}_{4}}^{\left(2\right)}$
is due to the nature of streamfunction vector. This may lead to different
consequences for interactions of directional waves.

\section{Summary\label{sec: sum}}

The original formulation of the $\boldsymbol{\Psi}$-formalism \citep{Kim01,KimBai03}
was derived under the constraints of irrotational flows and a homogeneous
bottom topography. This study investigates the consequences of removing
these constraints and presents a theory based on the $\boldsymbol{\Psi}$-formalism.

One convenient feature of the classic potential ($\Phi$) formalism
is that the interior flow structure may be completely eliminated using
its equation of motion and boundary conditions on the surface. The
dynamics may be described by a Hamiltonian function written entirely
in terms of the surface variables, i.e., free-surface displacement
$\eta$ and its canonical conjugate, the surface velocity potential
$\phi$. In reformulating the $\boldsymbol{\Psi}$-formalism, the
approach for developing the $\Phi$-formalism is applied to derive
a new form of the Lagrangian first proposed by Kim et al. \citeyearpar{Kim01,KimBai03},
based only on surface variables. This allows for separating problems
related to interior approximations from the issue of expressing $\boldsymbol{\Psi}$
in terms of $\phi$ and $\eta$. The governing equations that result
from taking variations of the Lagrangian with respect to the unknown
functions are organized in terms of dynamical equations for the canonically-conjugate
pair $\left(\eta,\phi\right)$, a surface constraint equation that
provides a connection between $\boldsymbol{\Psi}$ and $\left(\eta,\phi\right)$,
and the interior equation (essentially a condition that cancels two
horizontal components of the vorticity). For horizontally homogeneous
flows over mild topography, the Helmholtz-Hodge decomposition allows
for an asymptotic solution for the constraint equation, thus providing
the basis for deriving a surface Hamiltonian and Lagrangian. Expanding
these in terms of $\eta$ around a stationary background, we conclude
that the $\boldsymbol{\Psi}$-formalism leads to a Hamiltonian identical
to the standard potential formalism \citep{Zakharov99} for terms
up to fourth order. Furthermore, this formalism yields a new quartic
Lagrangian written in directly observable quantities $\eta$ and $\partial_{t}\eta$.
The description allows wave to propagate across background flows with
vertical vorticity. Applications of this feature will be presented
in future work.

\section*{Appendix. The $\boldsymbol{\Psi}$ -- $\Phi$ duality for potential
flows}

A curl-free (zero-vorticity) flow $(\mathbf{u},w)$ is fully determined
by the potential function $\Phi(\mathbf{x},z,t)$ as
\begin{equation}
\mathbf{u}=\nabla\Phi,\quad\mbox{and}\quad w=\Phi_{z},\label{eq: Phi-u}
\end{equation}
or, using differential forms \citep[e.g., ][]{blue Rudin}, as 
\begin{equation}
d\Phi=u^{j}dx_{j}+wdz,\label{eq: phi 0-form}
\end{equation}
with $\Phi$ a 0-form. For such a flow, the streamfunction $\boldsymbol{\Psi}$
\eqref{eq: u - Psi} provides an alternative description. Because
both $\boldsymbol{\Psi}$ and $\Phi$ completely define the flow,
they determine each other up to a total derivative. For potential
flows, this relationship is analogous to the so-called electromagnetic
(EM) duality \citep{Olive97}. Note that this duality exists only
for potential flows, since the $\Phi$ does not completely determine
flows with non-zero vorticity. 

Based on $\Phi$, the EM duality defines an 3-vector $\Psi'$ as
\begin{equation}
*d\Phi=d\Psi',\label{eq: EM Duality}
\end{equation}
where the star denotes the Hodge dual. The above relation defines
the 1-form $\Psi'$ through its differential, hence up to a differential
of a 0-form. Since the differential operator $d$ satisfies $d^{2}=0$,
one can use the 0-form $\sigma$ to shift $\Psi'$ by: $\Psi'\rightarrow\Psi'+d\sigma$
(gauge invariance). Without loss of generality, this allows to set
the $z$-component $(\Psi')^{z}=0$, which we shall henceforth assume.
The 1-form $\Psi'$ is now a 2-vector. One may easily check that a
$90^{\circ}$ rotation in the $\mathbf{x}$ plane produces a 2-vector
$\boldsymbol{\Psi}$ that is precisely the streamfunction defined
by \eqref{eq: Psi - u}. 

As is the case with EM duality, a rigid constraint in one of the variables
may become a soft constraint with the other. By applying the differential
operator directly to Eq. \eqref{eq: EM Duality}, one sees that

\[
d*d\Phi=d^{2}\Psi'\quad\Rightarrow\quad\triangle\Phi=d^{2}\Psi'.
\]

The right hand side is zero as a mathematical identity whereas the
left hand side is zero as the result of a condition on $\Phi$. The
physical meaning of both sides is that the fluid is incompressible.
Similarly, by applying first the Hodge dual to both sides of \eqref{eq: EM Duality}
and then the differential operator, we obtain
\[
d^{2}\Phi=d*d\Psi'.
\]
This is the condition of zero vorticity and it gives the condition
under which a mapping from $\Psi$ into $\Phi$ exists. As before,
a constraint for one variable, $\Phi$, becomes a Laplace equation
for the dual variable, $\Psi'$. 

In the $\boldsymbol{\Psi}$-formalism, the right hand side of the
above equation is not required to hold. Recall that we first use the
gauge invariance to set $(\Psi')^{z}=0$. If we take the variation
of the kinetic term $\frac{1}{2}\left(\mathbf{u}^{2}+w^{2}\right)$
in the usual definition of the action, without including the term
coming from varying $(\Psi')^{z}$, instead of $d*d\Psi'=0$, we simply
get
\[
\nabla(\nabla\cdot\Psi)+\partial_{z}^{2}\Psi=0.
\]
This is, therefore, a restricted form of `$d*d\Psi'=0$' which allows
for one component of vorticity. As a final note this form of EM duality
is non-local, meaning that $\Phi$ evaluated at a point $(\mathbf{x},z)$
may not be determined from $\boldsymbol{\Psi}$ and its derivatives
at that point alone, instead they are related through an integro-differential
transformation. Furthermore, this transformation generally exchanges
Neumann and Dirichlet boundary conditions, as we have already seen.

\section*{Acknowledgments}

This research was supported by the Office of Naval Research grants
N00014-10-1-0805 and N00014-10-1-0389.

\end{document}